\documentclass[manuscript]{acmart}
\AtBeginDocument{%
  \providecommand\BibTeX{{%
    \normalfont B\kern-0.5em{\scshape i\kern-0.25em b}\kern-0.8em\TeX}}}

\setcopyright{acmcopyright}
\copyrightyear{2018}
\acmYear{2018}
\acmDOI{XXXXXXX.XXXXXXX}

\acmConference[Conference acronym 'XX]{Make sure to enter the correct
  conference title from your rights confirmation emai}{June 03--05,
  2018}{Woodstock, NY}
\acmBooktitle{Woodstock '18: ACM Symposium on Neural Gaze Detection,
 June 03--05, 2018, Woodstock, NY} 
\acmPrice{15.00}
\acmISBN{978-1-4503-XXXX-X/18/06}

\usepackage{multirow}
\usepackage{booktabs}

\usepackage{float}
\usepackage{graphicx} 
\usepackage{caption} 
\usepackage{subcaption} 

\begin{document}


\title[The Interconnected Nature of Online Harm and Moderation]{The Interconnected Nature of Online Harm and Moderation: Investigating the Cross-Platform Spread of Harmful Content between YouTube and Twitter}

\author{Valerio La Gatta}
\orcid{1234-5678-9012}
\affiliation{%
  \institution{Information Sciences Institute, University of Southern California}
  \country{Los Angeles, California, USA}
}
\affiliation{%
  \institution{University of Naples Federico II}
  \city{Naples}
  \country{Italy}
}

\email{valerio.lagatta@unina.it}

\author{Luca Luceri}
\affiliation{%
  \institution{Information Sciences Institute, University of Southern California}
  \city{Los Angeles}
  \state{California}
  \country{USA}}
\email{lluceri@isi.edu}

\author{Francesco Fabbri}
\affiliation{%
  \institution{Pompeu Fabra University}
  \city{Barcelona}
  \country{Spain}}
\email{francesco.fabbri@eurecat.org}

\author{Emilio Ferrara}
\affiliation{%
  \institution{Information Sciences Institute, University of Southern California}
  \city{Los Angeles}
  \state{California}
  \country{USA}}
\email{ferrarae@isi.edu}

\renewcommand{\shortauthors}{La Gatta, et al.}

\begin{abstract}
The proliferation of harmful content shared online poses a threat to online information integrity and the integrity of discussion across platforms. Despite various moderation interventions adopted by social media platforms, researchers and policymakers are calling for holistic solutions. This study explores how a \textit{target platform} could leverage content that has been deemed harmful on a \textit{source platform} by investigating the behavior and characteristics of Twitter users responsible for sharing moderated YouTube videos. Using a large-scale dataset of 600M tweets related to the 2020 U.S. election, we find that moderated Youtube videos are extensively shared on Twitter and that users who share these videos also endorse extreme and conspiratorial ideologies. A fraction of these users are eventually suspended by Twitter, but they do not appear to be involved in state-backed information operations. The findings of this study highlight the complex and interconnected nature of harmful cross-platform information diffusion, raising the need for cross-platform moderation strategies.

\end{abstract}

\ccsdesc[500]{Information systems~Social networks}
\ccsdesc[300]{Human-centered computing~Empirical studies in collaborative and social computing}
\keywords{moderation interventions, cross-platform information diffusion, Twitter, YouTube}


\maketitle

\section{Introduction}


Social media platforms play a significant role in shaping the modern digital information ecosystem by allowing users to contribute to discussions on a wide range of topics, including public health, information technology, and socio-political issues. However, the freedom of expression offered by these platforms, combined with lax moderation policies, can potentially threaten the integrity of these information ecosystems when harmful content, such as fake news, propaganda and inappropriate or violent content, is shared and propagates across the digital population.
Mainstream social media platforms like Facebook and Twitter attempt to preserve the integrity of their environments by enforcing conduct policies and deploying various moderation interventions to target both harmful content and the users responsible for spreading it. These interventions can include flagging, demotion, or deletion of content, as well as a temporary or permanent suspension of users.



However, these moderation efforts are typically enacted in a siloed fashion, largely overlooking other platforms' interventions on harmful content that has migrated to their spaces. This approach poses risks as any inappropriate content originating on a \textit{source platform} can migrate to other \textit{target platforms}, gaining traction with specific communities and reaching a wider audience. For example, cross-platform diffusion of anti-vaccine content on YouTube and Twitter has led to extensive amplification and virality on multiple platforms \cite{10.1080/21645515.2021.2003647, 10.1038/s41598-020-73510-5}. Additionally, research has shown that moderation efforts on a source platform can foster the proliferation of harmful content on target platforms \citep{https://doi.org/10.48550/arxiv.2209.09803}. For example, the removal of anti-vaccine groups on Facebook has been found to increase engagement with anti-vaccine content on Twitter \citep{10.1145/3501247.3531548}. Similarly, \citet{10.1145/3449085} found that when YouTube decided to demote conspiratorial content, some Reddit communities pushed demoted videos on the platform making them go viral, effectively nullifying YouTube's strategy. Futhermore, \citet{10.1145/3447535.3462637} found that users who got banned on Twitter or Reddit exhibit an increased level of toxicity on Gab. 


Overall, this highlights the need for cross-platform moderation strategies that consider the interconnected nature of the digital information ecosystem, and that the removal of content or suspension of users on one platform may not be sufficient in addressing the spread of inappropriate content across multiple platforms. Cooperation among social media platforms is therefore desirable but also practically valuable: knowing what content has been deemed inappropriate on another platform can inform moderation strategies, or help with the early detection of similarly harmful, or related content. Recent work demonstrated how cross-platform strategies can help with the moderation of radical content or inauthentic activities \cite{DBLP:journals/corr/abs-1908-08313,10.1145/3485447.3512143}, e.g., by tracking users' activity on multiple platforms \citep{10.1145/3447535.3462637,10.1145/3501247.3531548, 10.1038/s41598-020-73510-5}.

\subsection*{Contributions of this work}
In this paper, we approach this problem from a different perspective. We consider YouTube (YT) and Twitter as the \emph{source} and \emph{target} platforms, respectively, and investigate the prevalence of moderated YT videos on Twitter, i.e., videos that are shared on Twitter but are eventually removed from YouTube. Also, we characterize Twitter users responsible for sharing YT videos---hereafter, YouTube \emph{mobilizers}, as defined by \cite{10.1145/3501247.3531571}---across several dimensions, including their political ideology and potential engagement with \emph{fringe} platforms.

In particular, we aim to answer the following research questions (RQs):


\begin{itemize}
     \item[\textbf{RQ1:}] \emph{What is the prevalence, lifespan, and reach of moderated YT videos that are shared on Twitter?} 
\item[\textbf{RQ2:}] \emph{What are the characteristics of the mobilizers of moderated YT videos? And, are there any differences with the mobilizers of non-moderated YT videos?}
\item[\textbf{RQ3:}] \emph{Do the mobilizers of moderated YT videos receive significant engagement from the Twitter population?} 
\end{itemize}

Leveraging a large-scale dataset related to the 2020 U.S. election \citep{DBLP:journals/corr/abs-2010-00600}, we observed that YouTube is the most shared mainstream social media platform on Twitter. By using the YouTube API to retrieve videos' metadata, we found that 24.7\% of the videos shared in the election discussion were moderated. We also found that these moderated videos spread significantly more than non-moderated videos and were shared more than content from other mainstream and fringe social media platforms, such as Gab and 4chan. When examining Twitter users sharing YT videos, we discovered that the users sharing moderated videos mostly engaged with YT content via retweets, while the users sharing non-moderated videos actively shared YT content in their original tweets or replies. We found that more than half of the users in the former group were suspended by Twitter, but surprisingly, there were more accounts involved in information operations---as identified by Twitter---in the latter group. Additionally, we found that the users sharing moderated YT videos supported Trump and promoted election fraud claims, while the users sharing non-moderated videos explicitly denounced Trump and had a more uniform political leaning, including users supporting both Biden and Republican representatives who did not endorse Trump's political campaign.

Finally, we found that users sharing moderated and non-moderated YT videos tend to interact within their group and have similar interaction patterns in terms of retweets, suggesting the formation of fragmented communities resembling echo chambers. Overall, our findings provide insights into the complex dynamics of cross-platform information diffusion, highlighting the need for a more holistic approach to moderation.

\section{Related Work}

\subsection{Cross-platform moderation}
To preserve the integrity of their own environment, most \emph{mainstream} social media platforms deploy diverse intervention strategies, targeting both inappropriate content (e.g., through flagging, demotion, or deletion) and the users (e.g., through temporary or permanent suspension) who share it. However, the effectiveness of these interventions is increasingly questioned by researchers and policymakers who demand for a proactive and holistic effort rather than the current siloed and retroactive solutions \citep{wilson2020cross, douek2020rise}. Indeed, even if the moderation intervention is effective on an isolated platform, 
it might trigger harmful activities on other platforms. 
For instance, \citet{10.1145/3447535.3462637} showed that, following the suspension of radical communities on Reddit, users migrated to alternative platforms becoming more active and sharing more toxic content. 
A similar pattern emerged following the deplatforming of Parler, where users migrated to other fringe social media platforms such as Gab and Rumble \cite{10.1093/pnasnexus/pgad035}. In addition, \citet{https://doi.org/10.48550/arxiv.2209.09803} have found that the antisocial behaviors of migrated users may spill over onto the mainstream one through other (non-radical) users active across platforms. Accordingly, \citet{10.1145/3501247.3531548} discovered that when Facebook banned some anti-vaccine groups, the toxic content promoted by these groups resonated on Twitter. Also, during the 2020 U.S. election, videos removed from \emph{mainstream} platforms were republished on the (low-moderated) BitChute platform \citep{https://purl.stanford.edu/tr171zs0069}. 

The above-mentioned studies raise the need for proactive and collaborative moderation approaches to guarantee the integrity of the whole digital information ecosystem. In this paper, we investigate the potential benefits 
of social media platforms to share information concerning their moderation interventions by studying the users who post moderated YT videos on Twitter.

 


\subsection{Cross-platform spread of YouTube content}


 Cross-platform multimodal information (e.g., images, videos) diffusion is particularly threatening as multimedia content is proven to be much more attractive and credible than only-textual posts \citep{doi:10.1080/10584609.2019.1674979}. In particular, the cross-posting of (harmful) video content across multiple social media platforms is a well-documented problem in the scientific literature \cite{wilson2020cross,doi:10.1177/1940161220912682}. For example, a considerable number of suspicious YT videos were shared on Twitter to raise skepticism regarding the COVID-19 vaccination campaign \citep{https://doi.org/10.48550/arxiv.2209.01675}. \citet{10.1038/s41598-020-73510-5} show that anti-vaccine YT videos that were shared on Twitter experienced an increased level of visibility and dissemination on YouTube, whereas \citet{10.1145/3501247.3531573} recognized YouTube as one of the most prolific channels used by the infamous \emph{Disinformation Dozen} to spread Covid-19-related conspiracies on Twitter. Similarly, \citet{doi:10.1177/1940161220912682} have shown that the Internet Research Agency (IRA) leveraged content from YouTube in their 2016 propaganda campaign on Twitter, and, more recently, \citet{wilson2020cross} have also reported the adoption of YT content to support anti-White Helmet operations in 2020. 
  

Overall, the above-mentioned studies demonstrate that YT harmful content is not only engaged within the \emph{source} platform but can flourish across other \emph{target} platforms, often with the intent of influencing vulnerable and fringe communities. In such a scenario,  the entities who mobilize and disseminate harmful content can take on various forms, i.e., bots, sockpuppets, influential elites, or even information consumers susceptible to misinformation and conspiracies. In this paper, we study the characteristics of these entities focusing on Twitter users who share moderated YT videos and investigating their similarities and differences with the users who share non-moderated YT content.

\section{Methodology}
In this section, we describe the data employed in the analysis and we detail the methodology used to understand the prevalence of YT moderated content on Twitter (\emph{RQ1}) and to characterize the users sharing moderated YT videos (\emph{RQ2} and \emph{RQ3}).  

\subsection{Data Collection}

\begin{figure}
\centering
\includegraphics[width=.75\columnwidth]{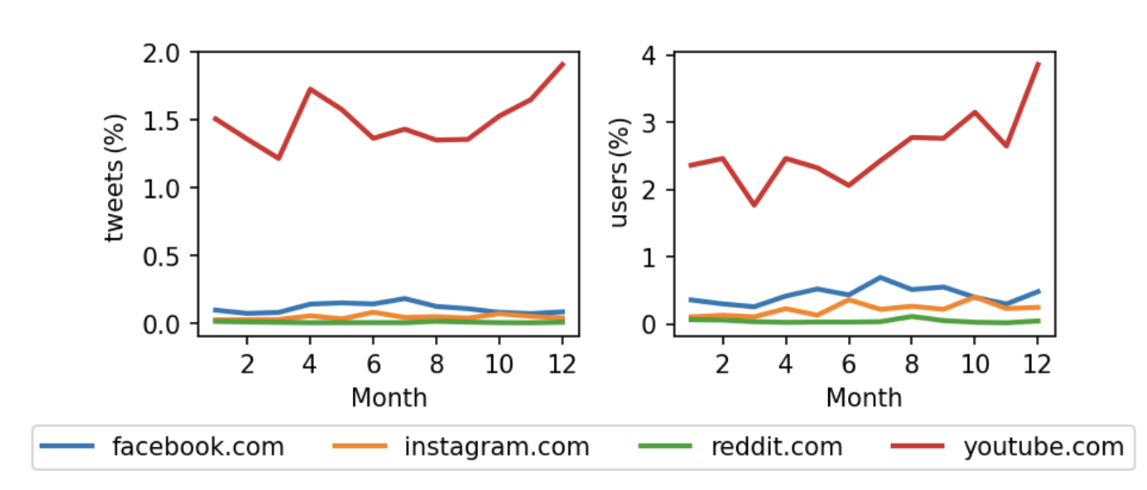} 
\caption{The monthly percentage of tweets (left) and users (right) who shared a link to a \emph{mainstream} social media platforms in 2020}
\label{fig:youtube_mainstream_trend}
\end{figure}

We employ a dataset of election-related tweets collected using Twitter’s streaming API service in the run-up to the 2020 U.S. election \citep{DBLP:journals/corr/abs-2010-00600}. In particular, we focus on the six months, from June 2020 to December 2020, covering the last part of the electoral campaign as well as the aftermath of the election, which was characterized by the widespread diffusion of misleading claims and conspiracies on the integrity of the election results \cite{suresh2023tracking,ferrara2020characterizing}. In this period of observation, we collect more than 600M tweets (including original tweets, replies, retweets, and quotes) shared by 7.5M unique users. In particular, tweets including YT videos account for the $0.65\%$ (3.9M) of the collected messages. Note that we do not consider URLs to YT channels. Fig. \ref{fig:youtube_mainstream_trend} shows that the fractions of tweets (resp. users) sharing YT content are consistently larger than tweets (resp. users) pointing  to other mainstream social media, which is in line with \cite{10.1145/3501247.3531571, Abilov_Hua_Matatov_Amir_Naaman_2021}.  In addition, in both cases, we observe an increasing trend towards our observation period (the second half of 2020) possibly because of the approaching election (November 3rd, 2020).


In total, 527k YT videos were shared on Twitter by 830k users. Through the YouTube API, we could retrieve various video metadata, which includes the ability to determine if a particular video was removed from the platform. We find that, among all YT videos, 24.7\% (130k out of 527k) were moderated. Besides, before the intervention,  those videos were shared on Twitter by 34.5\% (287k out of 830k) users. Finally, we collect YT video metadata of non-moderated videos, including the video title, description, tags, and the channel  that published the video. Interestingly, YouTube does not allow to collect metadata for moderated videos, including the date when the intervention occurred as well as the reason(s) for the moderation intervention. However, we can safely assume that a video is still online when shared in an original tweet as this requires the user to report the video URL in the Twitter post. 



\subsection{Identifying mobilizers of moderated YouTube videos}

To identify the mobilizers of moderated YT videos, we first consider the most active YT mobilizers, as we aim to 
investigate the characteristics and behaviors of users who repeatedly (rather than occasionally) post YT content. For this reason, we consider users that shared at least 5 YT videos on Twitter, which results in a set of 113k Twitter users. 

Then, we partition this list of YT mobilizers into two groups based on the volume of shared \emph{moderated} YT videos.
 
\begin{figure}[t]
     \centering

     \subfloat[][]{\includegraphics[width=.4\columnwidth]{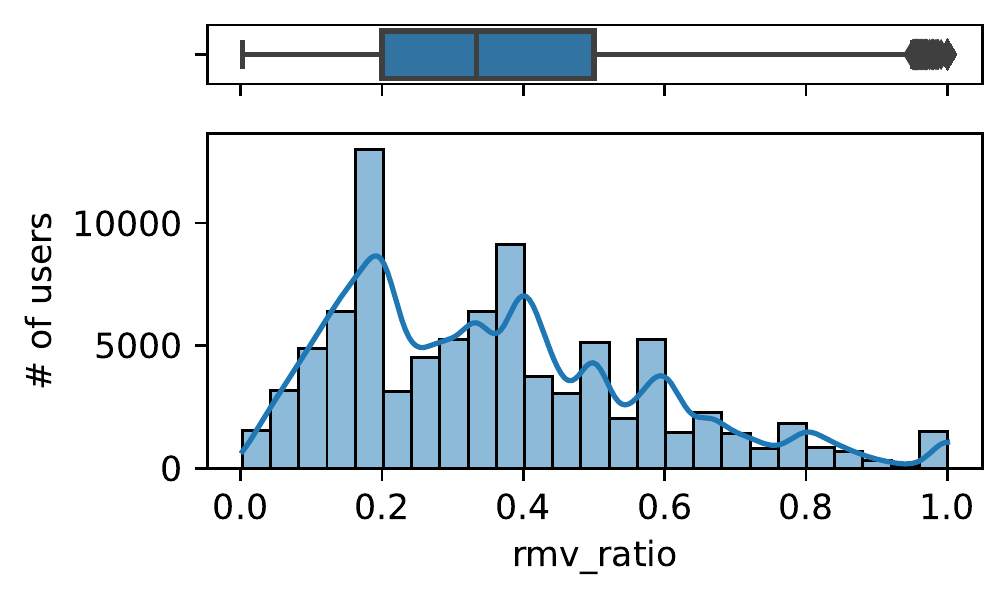}}
     \subfloat[][]{\includegraphics[width=0.4\columnwidth]{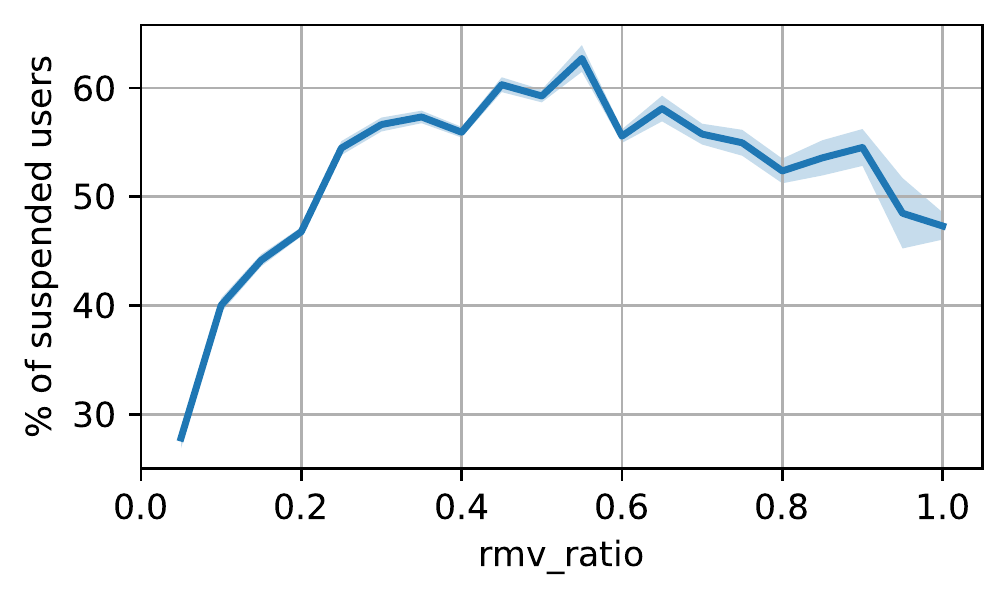}}
     
     \caption{YT Mobiliziers characteristics with: (a) Distribution of the \emph{rmv(u)}; (b) Percentage of the suspended users with respect to their \emph{rmv(u)}. The shaded area is the 95\% confidence interval.}
     \label{fig:rmv_ratio}
\end{figure}

For every user $u$, we define the \emph{ratio of moderated videos} (\emph{rmv(u)}) as the proportion of moderated YT videos out of the total number of YT videos shared by \emph{u} during our observation period. Based on this metric, we define non-moderated YT videos mobilizers (\textbf{NMYT}) as the users with an $rmv(u)=0$, which results in 25.4k Twitter accounts. 
Then, observing the distribution of the \emph{ratio of moderated videos} (Fig. \ref{fig:rmv_ratio}a), we define the mobilizers of moderated YT videos (\textbf{MYT}) the users with a $rmv(u) \geq 0.5$, i.e., all the users with a $rmv(u)$ higher than the 75-th percentile of the distribution, which results in 14.5k Twitter accounts. This threshold allows us to focus on users particularly prone to sharing moderated videos, thus excluding from the analysis the ones sporadically sharing moderated YT content.

To validate our choice, we examine whether the accounts sharing moderated videos are still active on Twitter or were suspended. Fig. \ref{fig:rmv_ratio}b shows the percentage of suspended users as a function of the \emph{rmv(u)}. We note that the percentage of suspended users does not increase after they share more than $50\%$ of moderated videos. Also, the probability of being suspended by Twitter is positively correlated with the value of the \emph{rmv(u)} (Spearman correlation = $0.451$).

\section{Results}

\subsection{Prevalence of moderated YouTube videos on Twitter (RQ1)}

To answer RQ1, we perform an analysis of the consumption of YT videos on Twitter comparing moderated vs. non-moderated YT videos.
It is worth noting that moderated videos tend to be shared for a limited number of days (20 days on average), while non-moderated videos have a longer lifespan (50 days on average), likely due to YT moderation interventions. 
Therefore, to perform a fair comparison, we examine the number of tweets including the YT videos during the first week after its first share on Twitter.

\begin{figure}[t]
     \centering
     \subfloat[][]{\includegraphics[height=3cm]{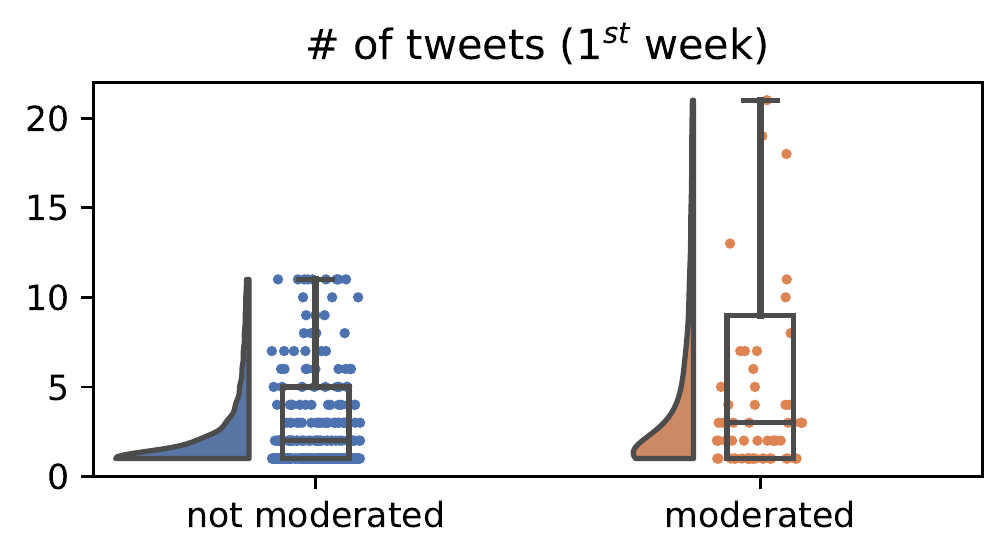}\label{fig:youtubeVSyoutubemod}}
     \subfloat[][]{\includegraphics[height=3cm]{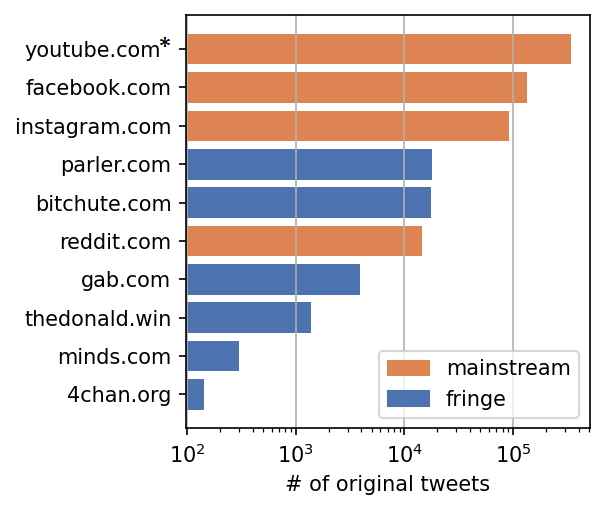}\label{fig:youtubeVSfringe_original}}
     \subfloat[][]{\includegraphics[height=3cm]{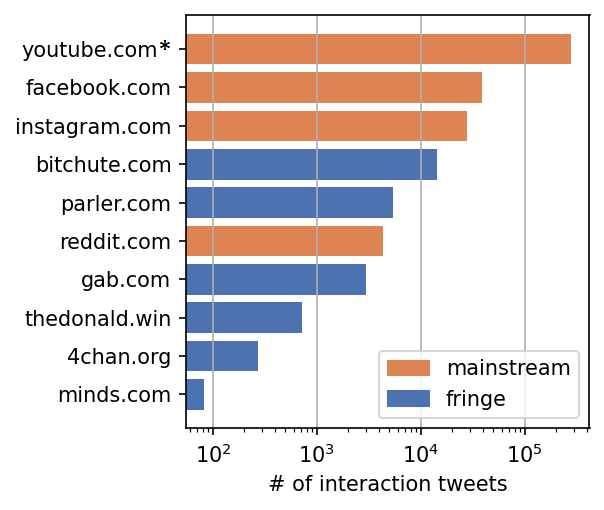}\label{fig:youtubeVSfringe_interactions}}
     \caption{ The prevalence of moderated YT videos with: (a) The distribution of the number of tweets sharing each video during the week after its first share; (b) Number of original tweets containing a link to each social media platform (Log-scale); (c) Number of retweets containing a link to each social media platform (Log-scale)}
     \label{youtubeVSfringe}
\end{figure}

Fig.~\ref{fig:youtubeVSyoutubemod} shows the distributions of the number of tweets for non-moderated and moderated videos. It can be noted how moderated videos distribution is characterized by a right heavy-tail, meaning that moderated videos when posted for the first time on Twitter, generate a higher volume of sharing activity with respect to non-moderated videos\footnote{A Mann–Whitney test ($p$-value$<0.01$) was performed to validate this finding.}. 

To further explore the prevalence of moderated YT content on Twitter, we compare the volume of interactions with content originating from other social media platforms. Specifically, Fig.~\ref{fig:youtubeVSfringe_original} and Fig.~\ref{fig:youtubeVSfringe_interactions} shows the volume of tweets and retweets of moderated YT videos, URLs pointing to \emph{mainstream} online social networks, and URLs redirecting to fringe platforms \citep{9671843}. 
%
We observe that the volume of tweets linking \emph{moderated} YT videos alone is greater than the volume of tweets pointing to any other social media platform. In line with \cite{10.1145/3501247.3531571}, we find that \emph{fringe} content supplied by \emph{Parler} and \emph{BitChute} is outnumbered by the content supplied by mainstream platforms.


\subsection{YouTube Mobilizers (RQ2)}

To address RQ2, 
we characterize YT mobilizers that share moderated (MYT) and non-moderated videos (NMYT) and investigate whether these users show significantly different behaviors and characteristics across three dimensions:

\begin{itemize}
    \item \emph{Cross-Posting Activity}: we explore the sharing activity that users perform on Twitter, including their cross-posting of content originating from other \emph{mainstream} and \emph{fringe} platforms;     
    \item \emph{Trustworthiness of the account}: we examine whether MYT and NMYT users are verified accounts or bots, also looking at their account status (active vs. suspended) and potential involvement in information operations;  
    \item \emph{User Interests}: we investigate the political leaning and topics of interest of mobilizers, both on Twitter and YT. 
\end{itemize}

\subsubsection{Cross-Posting.}

\begin{figure*}[t]
\centering
\includegraphics[width=0.9\textwidth]{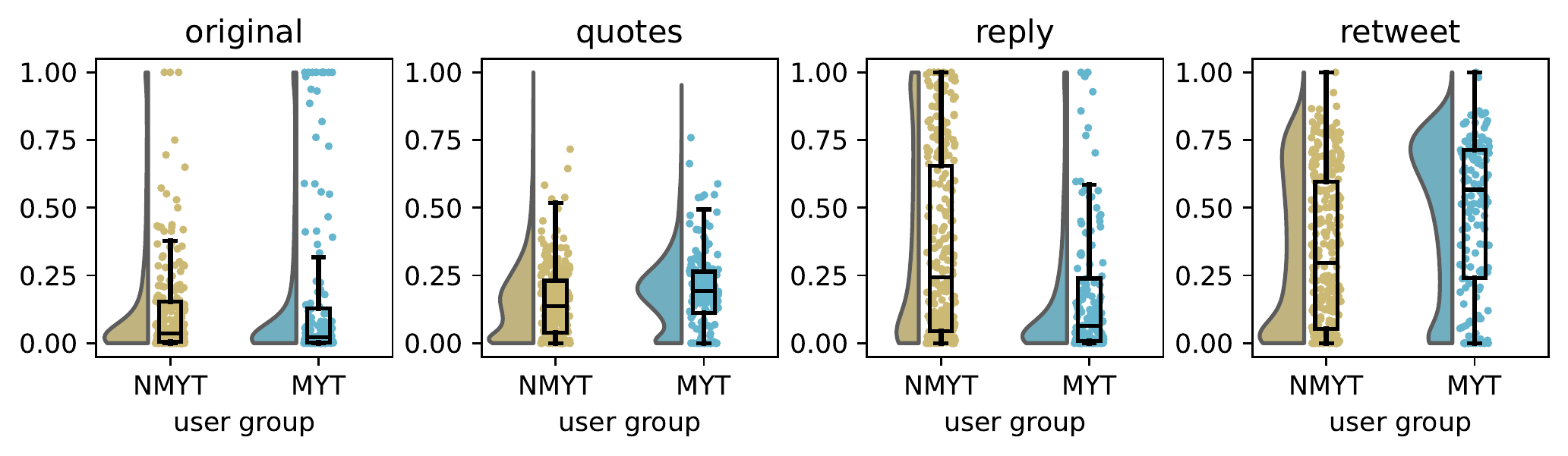} 
\caption{The distribution of original tweets, replies, retweets and quotes for NMYT and MYT mobilizers }
\label{fig:users_actions}
\end{figure*}

\begin{figure*}[t]
     \centering

     \subfloat[][]{\includegraphics[width=0.3\textwidth]{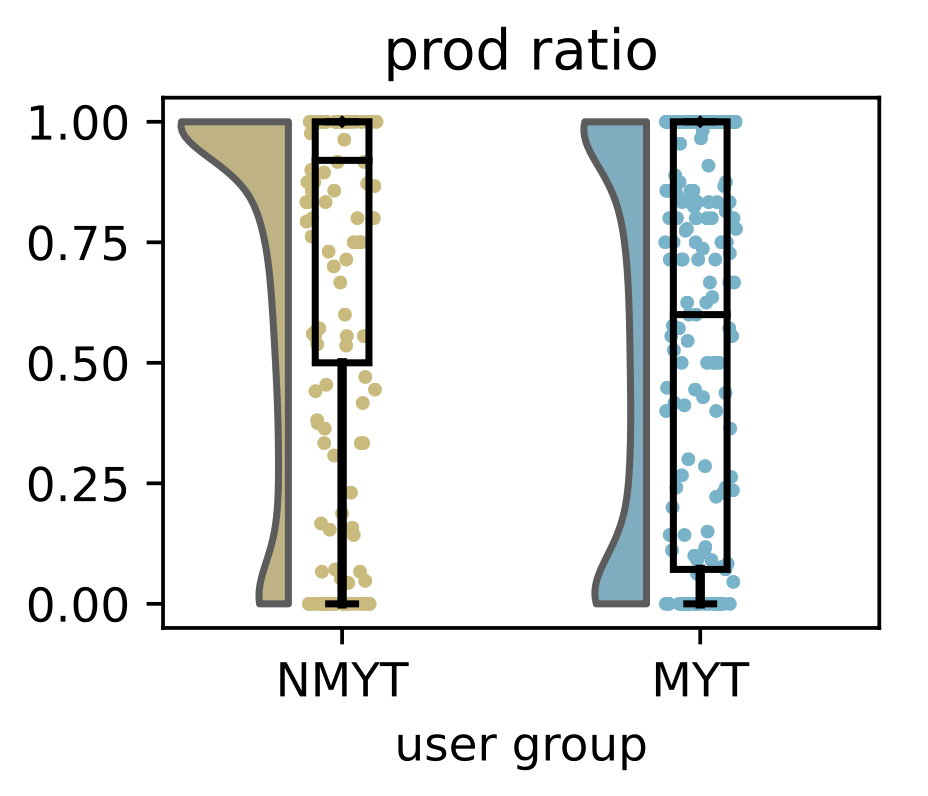}\label{fig:prod_ratio}}
     \subfloat[][]{\includegraphics[width=0.3\textwidth]{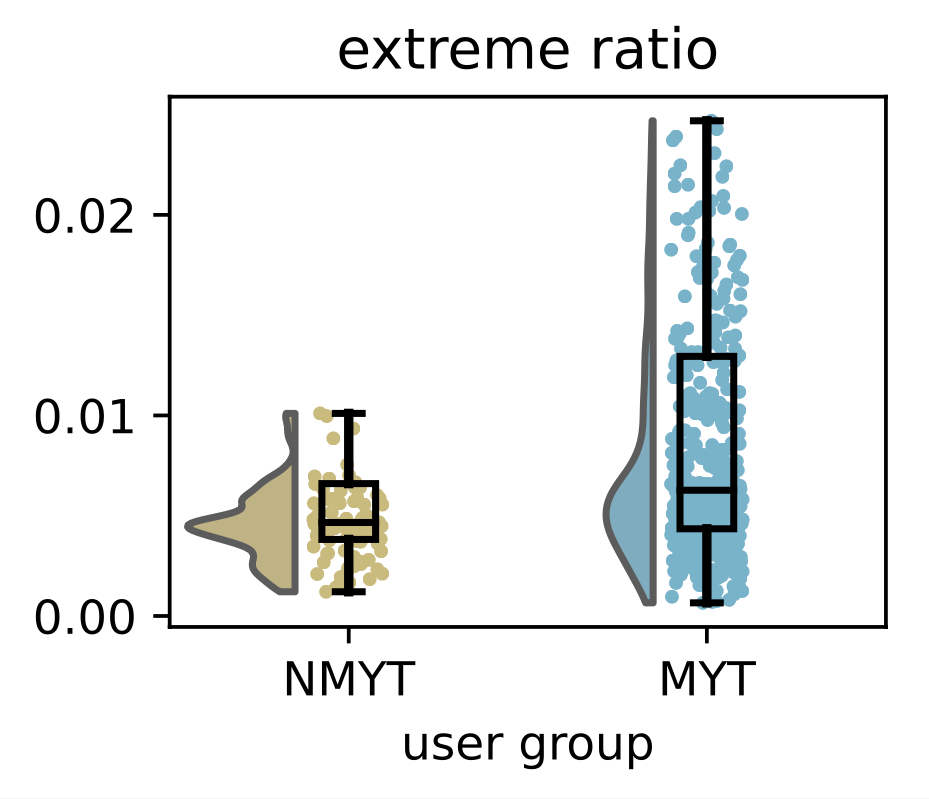}\label{fig:extreme_ratio}}
     \subfloat[][]{\includegraphics[width=0.29\textwidth]{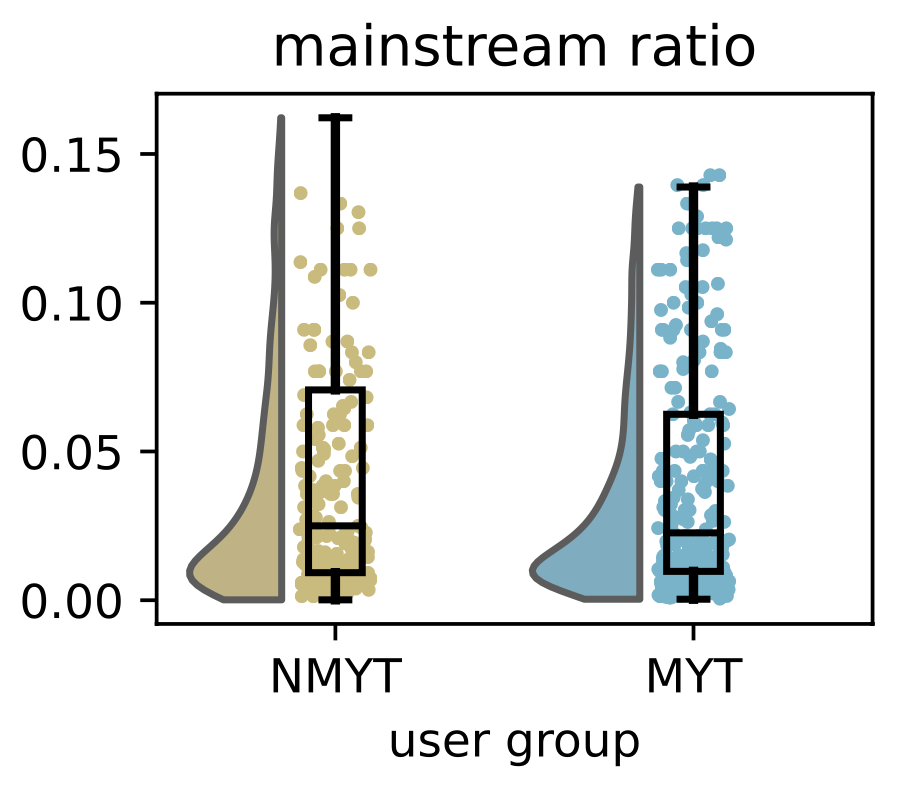}\label{fig:mainstream_ratio}}     
     \caption{The distribution of \emph{prod\_ratio}, \emph{extreme ratio} and \emph{mainstream ratio} for NMYT and MYT mobilizers}
     \label{fig:users_prod_ext}
\end{figure*}

We consider all the possible sharing activities that a user can perform on Twitter, i.e., posting an \emph{original} tweet, commenting on a tweet with a \emph{reply}, re-sharing a tweet with or without a comment via \emph{retweet} or a \emph{quote}, respectively. 
 
From Fig. \ref{fig:users_actions} it can be noticed that, while exhibiting similar behaviors in posting \emph{original} tweets, MYT users retweet more and reply less than NMYT users. To further characterize this discrepancy in the use of retweets, we examine the proportion of re-shared tweets embedding links to other web domains (beyond YouTube). We find that MYT users often retweet content linking to external sources, i.e., $50\%$ of their retweet on average contain URLs with respect to $28\%$ for NMYT mobilizers. This analysis suggests that NMYT users' activity is more diversified on Twitter while MYT users tend to passively consume and spread (through retweets) the content they see on Twitter, especially if such content links to external resources. 

To further characterize the two groups of mobilizers, we consider how they engage with YT videos on Twitter. Specifically, we aim at understanding whether users passively interact with YT videos (e.g., through retweets) or they pro-actively share them (e.g., in their original tweets). 
To this end, we define users as a \emph{producers} (resp. \emph{consumers}) of a YT video if their first share of that video is an original tweet or a reply (resp. retweet). Then, we define the \emph{prod\_ratio} as the proportion of YT videos \emph{produced} by a user out of the total number of videos that he/she engage with. It is worth to note that with ``video producer'' we do not imply that the user is the publisher of the video on YT. 

Interestingly, we find that the mobilizers in both groups are either mostly producers (\emph{prod\_ratio $>80\%$}),  or mostly consumers (\emph{prod\_ratio $<20\%$}) of YT videos. Specifically, we count 15,761 (62.1\%) producers and 4,439 (17.4\%) consumers in the NMYT group, while there are 
4,630 (31.9\%) producers and 3,742 (25.8\%) consumers belong to the MYT group.
Fig. \ref{fig:prod_ratio} shows the distribution of the \emph{prod\_ratio} in both groups. On the one hand, we find that NMYT mobilizers are mostly producers, i.e., the 25-percentile of the \emph{prod\_ratio} distribution is $0.50$. On the other hand, the MYT group includes the same amount of producers and consumers accounts. This result is consistent with our previous finding and confirms MYT users' tendency to passively retweet content if compared to NMYT accounts, who actively participate in the discussion through replies or quotes. 

Furthermore, we investigate how mobilizers engage with content from other social media platforms. In particular, we target \emph{mainstream} platforms, i.e., Facebook, Instagram, and Reddit, and the \emph{fringe} platforms outlined in Fig. \ref{fig:youtubeVSfringe_original}. We define the \emph{extreme ratio} 
as the fraction of extreme URLs 
out of the total number of tweets shared by each user. 

Fig. \ref{fig:mainstream_ratio} shows that  MYT and NMYT mobilizers have similar distributions in terms of \emph{mainstream ratio} ($p$-value $>0.01$ with a Mann-Whitney test). Altogether, MYT and NMYT mobilizers share the same percentage ($3.3\%$) of \emph{mainstream} content on average, and no user is posting more than $15\%$ of tweets linking other \emph{mainstream} social media platforms. However, Figure \ref{fig:extreme_ratio} shows that MYT users engages with \emph{fringe} platforms much more than NMYT mobilizers.  Indeed, the distribution of their \emph{extreme ratio} are different according to a Mann-Whitney test ($p$-value$<0.01$), even if they have similar mean values, i.e., $0.5\%$ and $0.7\%$ for NMYT and MYT mobilizers, respectively. Overall, this analysis highlights that the two groups of mobilizers do not exhibit any difference when interacting with \emph{mainstream} social media platforms, whereas MYT mobilizers share more content from low-moderated online spaces than NMYT users, suggesting a form of endorsement towards the extreme ideas pushed on \emph{fringe} platforms \citep{9671843}. 

\subsubsection{Trustworthiness.}

\begin{figure*}[t]
\centering
\includegraphics[width=0.9\textwidth]{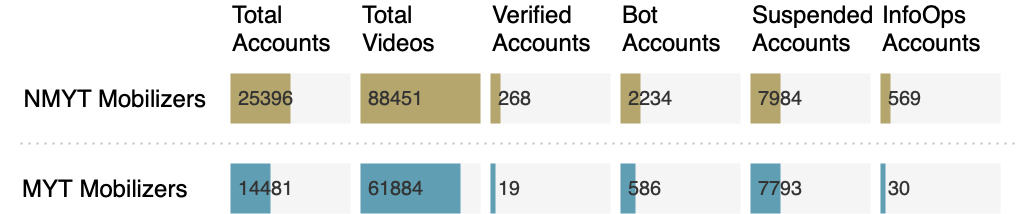} 
\caption{The number of accounts in each mobilizer group that were verified, bots or suspended. The columns are as follows: ``Total Accounts'' is the total number of accounts in each group. ``Total Videos'' is the number of unique YT videos shared by each group. ``Verified Accounts'' is the number of verified accounts in each group. ``Bot Accounts'' is the number of accounts labeled as a bot by the Botometer API in each group. ``Suspended Accounts'' is the number of accounts in each group that were later suspended by Twitter. ``InfoOps Accounts'' is the number of (suspended) accounts involved into information operation in each group}
\label{tab:stats}
\end{figure*}

Here, we investigate the nature and status of the accounts in the two groups of mobilizers. Given the pivotal role that political elites, bot accounts, state-backed trolls have in orchestrated campaigns and (mis)information operations \citep{luceri2020detecting, doi:10.1177/1940161220912682,doi:10.1080/1369118X.2019.1621921,10.1145/3501247.3531573},
we aim at identifying the entities pushing moderated or non-moderated YT videos on Twitter. To this end, we leverage Botometer \citep{DBLP:journals/corr/abs-2201-01608} and the Twitter API to assess whether our mobilizers are automated or verified accounts, respectively. As shown in Figure \ref{tab:stats}, we find that both groups include very few verified accounts ($268$ and $19$ in the NMYT and MYT groups, respectively) and bots ($2,234$ and $586$ in the NMYT and MYT groups, respectively). 

Next, we examine whether users in the two groups of mobilizers were suspended by means of Twitter moderation interventions. Indeed, during the 2020 U.S. election, the platform made an increased effort to guarantee the integrity of discussion by adding warnings to suspicious or misleading content, as well as suspending accounts involved into information operations \citep{luceri2021social,sanderson2021twitter}. From our perspective, we are interested in understanding whether and to what extent the accounts of the MYT and NMYT mobilizers were moderated by these actions. We find that accounts are suspended in both groups but in different proportion: Fig. \ref{tab:stats} shows that $53.8\%$ of MYT mobilizers (7,793 accounts) are moderated by Twitter, while $31.4\%$ of NMYT users (7,984 accounts) are suspended. In addition, we further explore whether the (suspended) accounts are involved in state-backed information operations (in short, InfoOps) on Twitter. 
As reported in Figure \ref{tab:stats}, we find that a minority of mobilizers (599 accounts in total) are involved in those campaigns and, interestingly, NMYT users are more involved than MYT users ($2.2\%$ of MYT against just $0.2\%$ of NMYT). Overall, this analysis suggests that even though MYT mobilizers violated Twitter policies, they are not involved in state-backed, orchestrated campaigns during the election. 


\subsubsection{User Interests.}

We now turn our attention to the content shared by MYT and NMYT mobilizers. As the data under analysis relate to the U.S. 2020 Presidential election, we expect that the discussion resonate political topics, especially regarding the electoral campaign of the candidates, as well as the heated discussion around the alleged evidence of fraud during the aftermath of the election. Also, we focus on the political orientation of the users under analysis, and on their potential connection with conspiratorial and fringe theories. 

To explore the general interests of MYT and NMYT mobilizers, we compare the hashtags of their tweets, as well as the descriptions of the YT videos they share on Twitter. 
To this aim, we apply SAGE \citep{10.5555/3104482.3104613} to find the most distinctive hashtags and keywords for the tweets and the video descriptions, respectively, shared by the two groups. It is worth to note that for MYT users we only took into account the non-moderated videos as the YouTube API cannot retrieve metadata related to moderated content. Table \ref{tab:keywords} reports the 
keywords and hashtags extracted by SAGE. We  observe that MYT users are Trump's supporters, as showed by the hashtags \emph{\#bestpresidentever45, \#demonrats}, and they also sustain his allegation of voter fraud after the electoral count, as noticeable by the hashtags \emph{\#krakenteam, \#trumpwon}. On the contrary, NMYT mobilizers explicitly despise Trump, as it can be noticed by the hashtags \emph{\#trumpvirus, \#traitorinchief}, but their political orientation is not as clear as for MYT users. Indeed, \emph{\#gojoe} is the only pro-Biden hashtag in the NMYT's top-50 hashtags. 

Overall, the keywords extracted from YT videos descriptions are aligned with the Twitter hashtags of the two groups. However, they do not communicate positive or negative sentiment but usually refer to (groups of) people who publicly stated their political preference. For instance, NMYT Mobilizers shared several videos related to the \emph{lincoln} project and its ad starring \emph{Barkhuff Dan} explicitly saying ``I can see Trump for what he is — a coward. We need to send this draft-dodger back to his golf courses''. It is worth to note that the Lincoln project is run by Republicans opposing Trump, which supports our intuition that NMYT users are not always Biden supporters. On the contrary, the MYT mobilizers support for Trump is clear also from their shared videos mentioning \emph{Christina Bobb}, who has been close to the Trump's legal team that tried to overturn the result of the Presidential election, and \emph{Tucker Carlson}, who has been recently nicknamed as Trump's heir\footnote{https://www.theguardian.com/media/2020/jul/12/tucker-carlson-trump-fox-news-republicans}. 

\begin{figure}
     \centering

     \subfloat[][]{\includegraphics[height=4cm]{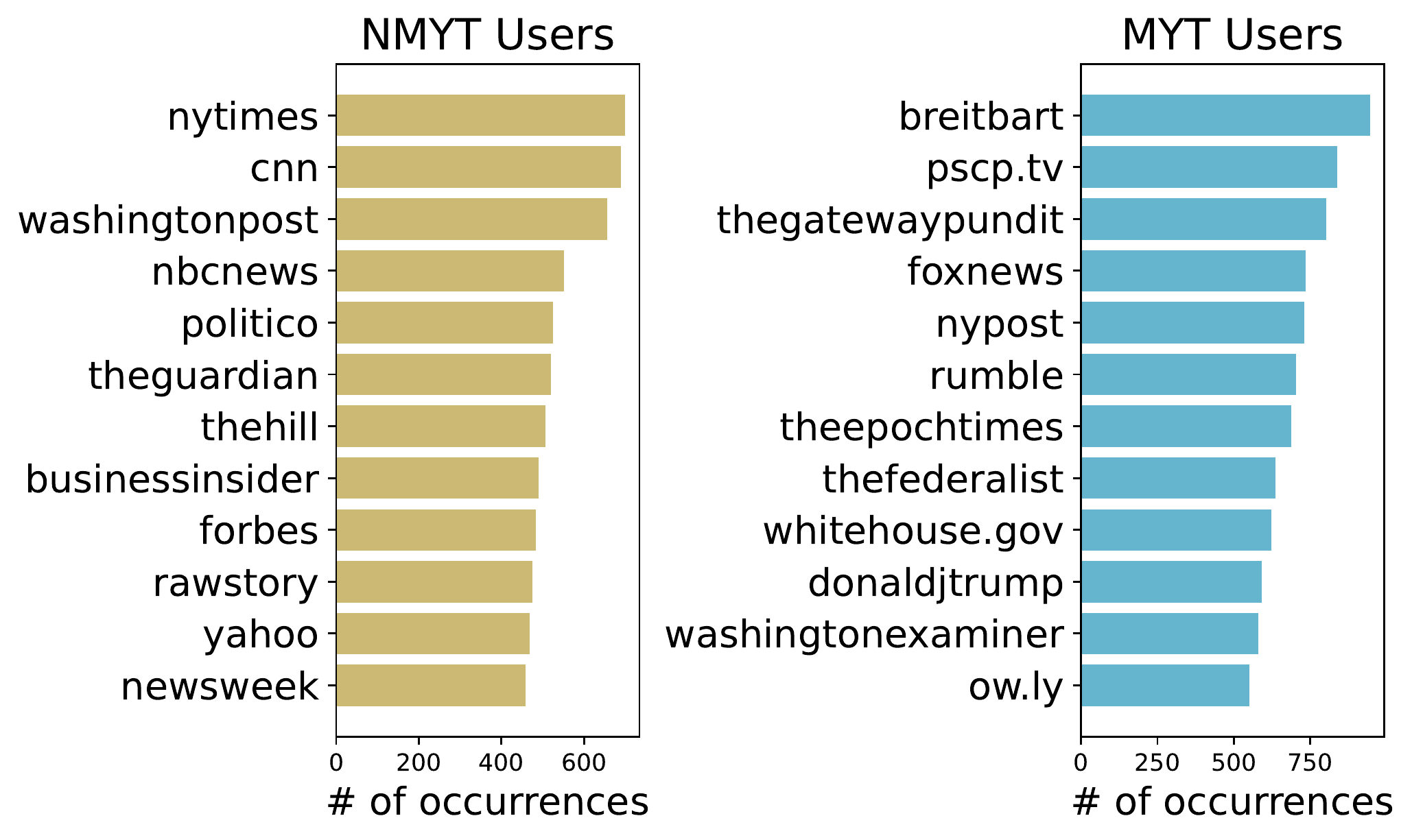}\label{fig:news_outlets}}
     \subfloat[][]{\includegraphics[height=4cm]{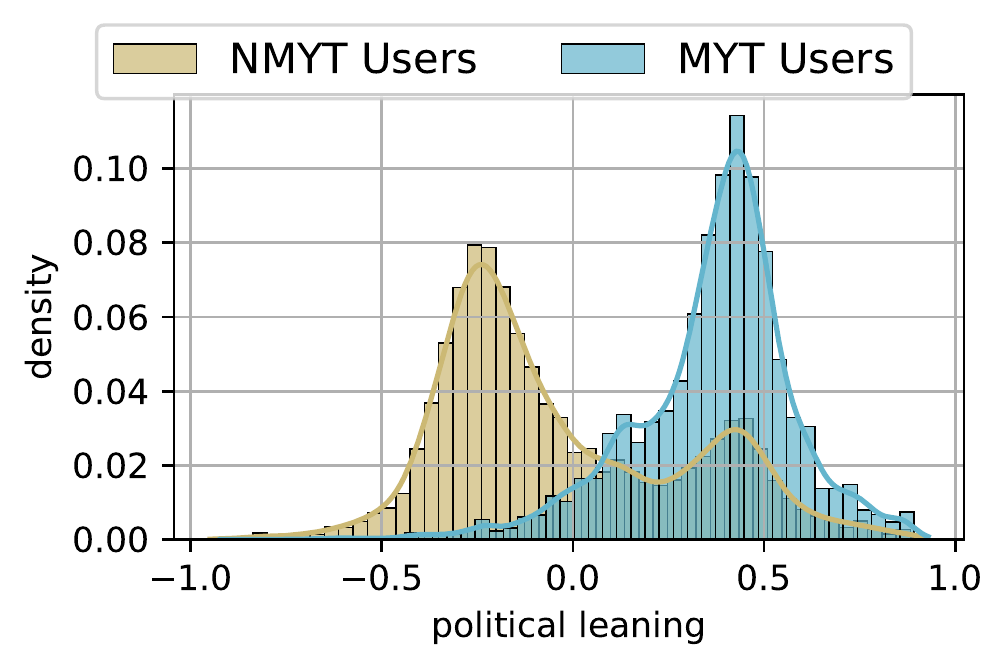}\label{fig:political_leaning}}
     
     \caption{(a) The news outlet shared by each group of mobilizers (we omit the \emph{.com} extension for brevity) ; (b) the distribution of the political leaning within the two groups of mobilizers }
     \label{political}
\end{figure}

To further investigate users' political orientation, we leverage the political leaning score assigned by MediaBiasFactCheck to several news outlets and, consistently with previous works \cite{Ferrara_Chang_Chen_Muric_Patel_2020,10.1145/3308560.3316494}, we measure users' political orientation by averaging the scores of the domains they share on Twitter during the observation period. Fig. \ref{fig:news_outlets} and \ref{fig:political_leaning} show the top-10 domains shared by YT mobilizers and the political leaning distribution of MYT and NMYT users. The former group includes several far-right users (the median of the distribution is $0.47$) who mostly share news from \emph{breitbart.com} and \emph{thegatewaypundit.com}, which are known to promote conspiracy theories and publish extreme conservative content. On the contrary, the latter group includes less extreme and more liberal users. However, the political leaning distribution of the NMYT group is bimodal (the larger mode is $-0.27$ and the smaller one is $0.41$) and a small subset of these mobilizers shows an extreme, conservative ideology. This result is further confirmed by looking at the top-10 domains of MYT mobilizers, which include \emph{foxnews.com}, which is also shared frequently by the NMYT group, and \emph{forbes.com}, which is a center-right news outlet.

\begin{table}
    \centering
    \caption{Most shared hashtags and YT video keywords 
    by NMYT and MYT mobilizers}
    \scalebox{.95}{
    \begin{tabular}{rcc}\toprule
    
  & \textbf{NMYT Mobilizers} & \textbf{MYT Mobilizers}  \\
 \midrule
\multirow{8}{*}{\begin{tabular}[c]{c@{}}\textbf{Twitter} \\ \textbf{hashtags} \end{tabular}} 
& \#putinspuppet & \#krakenteam \\ 
& \#trumpvirus & \#chinebitchbiden \\
& \#resignnowtrump & \#demonrats \\
& \#trumplies & \#evidenceoffraud \\ 
& \#traitorinchief & \#bestpresidentever45 \\
& \#gojoe & \#arrestfauci \\ 
& \#trumpkillus & \#trumpwon \\
& \#weirdotrump & \#trumppatriots \\  \midrule

\multirow{7}{*}{\begin{tabular}[l]{r@{}}\textbf{YouTube} \\ \textbf{keywords} \end{tabular}} 
& barkhuff dan & rsbn \\
& bernie sander & bobulinski \\
& lincoln & censored \\ 
& rainbow & christina bobb \\
& loyalty & fitton \\ 
& cnn &  spoiled \\
& incompetence & tucker carlson \\ \bottomrule
    
    \end{tabular}}

    \label{tab:keywords}
\end{table}

\section{Engagement towards mobilizers of moderated YouTube videos (RQ3)}

To respond to RQ3, we analyze the interaction patterns enacted by MYT and NMYT mobilizers by looking at the intra- and inter-group retweets exchanged by the users. In addition, we also consider the group of all \emph{other} users (750k Twitter accounts) who shared at least one YT video and, by definition, do not fall in the category of YT mobilizers. 

\begin{figure*}[t]
     \centering

     \subfloat[][]{\includegraphics[height=4.15cm]{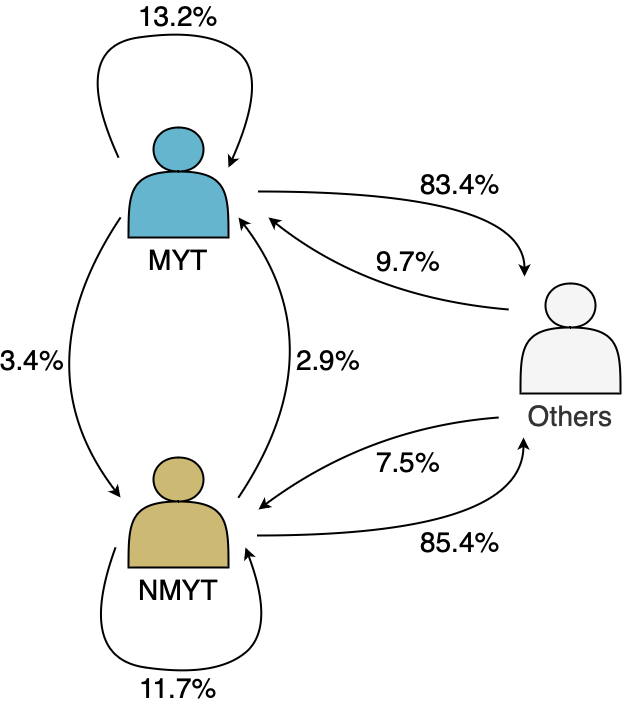}\label{fig:rt_src}}\hspace{5mm}
     \subfloat[][]{\includegraphics[height=4.15cm]{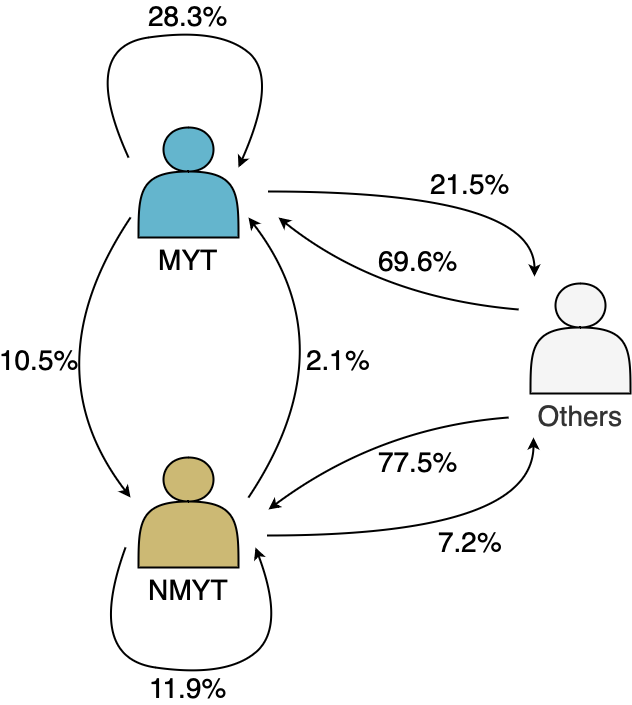}\label{fig:rt_dst}}\vspace{1mm}
     \subfloat[][]{\includegraphics[height=4.15cm]{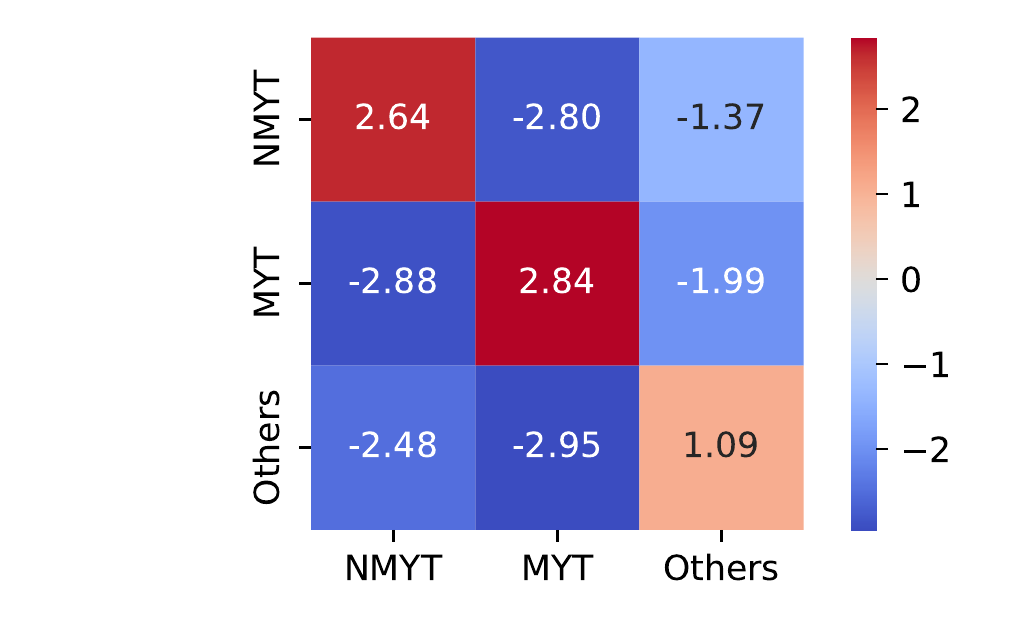}\label{fig:nullmodel}}
     
     \caption{Interaction patterns enacted by NMYT and MYT accounts: (a) Proportion of interactions between YouTube mobilizers normalized by the source; (b) Proportion of interactions between YouTube mobilizers normalized by the destination; (c) Z-scores of observed retweets between YouTube mobilizers ($p$-value $< 0.01$) }
     \label{ret_net}
\end{figure*}

As the number of users in the three groups is different, we do not compare the absolute numbers of intra- and inter-group retweets. For this reason, we normalize the number of interactions by source (i.e., the total number of retweets that the group performs, see Fig. \ref{fig:rt_src}) or by target (i.e., the total number of retweets that the group receives, see Fig. \ref{fig:rt_dst}). 
On the one hand, we observe that MYT and NMYT mobilizers generate the same relative amount of intra-group retweets, $13.2\%$ and $11.7\%$, respectively, and inter-group retweets, $3.4\%$ and $2.9\%$, respectively. On the other hand, Fig. \ref{fig:rt_dst} highlights that the two groups are engaged differently when considering the proportion of retweets they receive, i.e., $28.3\%$ of the retweets received by MYT users are from accounts of the same group and only $2.1\%$ are from the NMYT mobilizers. However, NMYT mobilizers tend to retweet both groups with the same frequency  ($11.9\%$ NMYT to NMYT and $10.5\%$ MYT to NMYT). This result indicates that MYT users tend to retweet and be retweeted more within their group than between groups.
Finally, we can also notice that the users in the \emph{Others} category retweets MYT and NMYT mobilizers almost in the same way. This observation emphasizes that the level of user activity is not informative to characterize interactions between NMYT users and the rest. 

To further validate our results, we compare the observed number of retweets to a null model, which assumes that interactions occur by chance. Specifically, we randomize the users' assignment to the three groups (i.e., NMYT, MYT, and Others) and compute the mean and standard deviation of the interactions between the groups for 100 iterations. We then compute the z-scores to compare the observed retweets with the expected number of retweets from the null model. Fig. \ref{fig:nullmodel} indicates that the observed pattern of retweeting behavior among the mobilizers is consistent with the principle of homophily. Specifically, both groups of mobilizers were found to have a greater number of retweets within their respective groups and a lower number of retweets across groups than what would be expected by chance, i.e., $z > 2.5$ and  $z < -1.5$, respectively.



\section{Discussions}

\subsection{Contributions}
In this paper, we studied the Twitter discussion around the (video) content that is deemed harmful on YouTube. Leveraging an unprecedented large-scale dataset of 600M tweets shared by over 7.5M users, we discovered an unexpectedly high number of \emph{moderated} YT videos shared on Twitter during the run-up to and aftermath of the 2020 U.S. election. Overall, moderated videos were shared more than non-moderated ones and received far more attention than content from \emph{fringe} social media platforms. Moving beyond previous work, we investigated the characteristics of the Twitter users responsible for sharing both moderated and non-moderated YT videos. On the one hand, we found that users sharing moderated content tend to passively retweet what they see on Twitter rather than actively posting original tweets or replies. On the other hand, the mobilizers of non-moderated videos actively share YT videos in their original tweets. Overall, most of the users were regular Twitter accounts rather than bots or state-sponsored actors, and, even if we did not find any involvement in information operations, more than half of the mobilizers of moderated videos were suspended by Twitter. Furthermore,  we found that the mobilizers of moderated YT videos are far-right supporters and sustained Trump during the 2020 U.S. election. By contrast, the political preference of the mobilizers of non-moderated YT videos is more diverse since users in this group range from Biden supporters to other republican representatives who did not endorse Trump's political campaign. Finally, we studied the interactions between the mobilizers of moderated and non-moderated videos and discovered that both groups exhibit strong group cohesion and are engaged by the general Twitter audience in a similar manner.

\subsection{Limitations}
There is a number of limitations in our study. First, neither Twitter nor YouTube provides any supplementary information about account suspension and video moderation, as well as we do not know when these interventions occurred. As a result, there is no guarantee that YT videos were still online when re-shared through retweets on Twitter, but we can confidently assume they were not moderated yet when shared in an original tweet. In addition, we acknowledge our analyses, like several previous works \citep{10.1145/3511095.3531283,9381310}, might be biased towards moderated YT content which includes not only videos violating YouTube policies but also those removed by their publishers for any reason. 

Second, we overlooked YT channels shared on Twitter to safeguard our analysis from Twitter users who just advertise their own (or others) YouTube channel \citep{10.1145/2556195.2566588}. However, this choice might prevent us to consider another potential source of harmful YT content on Twitter. 

Third, the partition strategy to define the two groups of mobilizers is quite conservative as we considered Twitter users who \emph{never share} moderated videos (NMYT mobilizers) and those who \emph{mostly share} moderated videos (MYT mobilizers). The investigation of the users sharing a limited number of moderated videos is left for future research.

\subsection{Conclusions and Future Works}

Our study has two major takeaways: first, moderated YT videos are extensively shared on Twitter, and users who (passively) share those, endorse extreme and conspiratorial ideas. From a broader perspective, we have shown how harmful content originating in a \emph{source} platform significantly pollute discussion on a \emph{target} platform. Although more research is still needed, we conjecture that sharing information about the taken interventions would improve our understanding of cross-platform harmful content diffusion and benefit all entities within the information ecosystem. For instance, in the YouTube-Twitter cross-posting scenario considered in this paper, YouTube moderation activity can benefit both parties of the cooperation: on the one hand, Twitter has the opportunity to (early-)detect intra-platform harmful activities; on the other hand, YouTube can further improve its moderation based on the cross-platform signals tied with harmful YT content diffusion on Twitter.


Second, the mobilizers of moderated YT videos were rarely  bot accounts or involved in information operations, but instead appeared to be regular Twitter users who do not necessarily share content from \emph{fringe} platforms. This suggests that the cross-posting of (harmful) cross-platform content is participatory \cite{luceri2021social} and research in this field should not only target bots and trolls but, instead, consider the role of online crowds and more complex social structures across different social media platforms.

Future work might build upon our findings to design algorithms to automatically identify or predict whether a YT video will be moderated based on the engagement it receives on Twitter, as well as detect early-signals of radicalization. In addition, we aim to investigate whether our results generalize to other highly-moderated social media (e.g. Facebook, Instagram), including case studies considering content originating from low-moderation, fringe spaces.



\bibliographystyle{ACM-Reference-Format}
\bibliography{sample-base}

\end{document}